\documentclass[aps,prd,preprint
]{revtex4-2}
\usepackage{graphicx} 

\usepackage{amsmath}
\usepackage{tensor}
\usepackage{mathrsfs}
\usepackage{amsfonts}
\begin{document}
\title{Complex Quaternionic Formulations of Dirac, Electrodynamic, and Electroweak Fields and Interactions}
\author{James Henry Atwater, David Lambert, Yuri Rostovtsev}
\affiliation{Center for Nonlinear Sciences and Department of Physics, University of North Texas, Denton, TX 76203, USA}

\date{February 2026}

\begin{abstract}
    
A simple translation between a standard representation of $\mathfrak{sl}_2\mathbb{C}$ and the complex-quaternions ($\mathbb{H}\otimes_\mathbb{R}\mathbb{C}$) is established and exploited to construct a novel hyper-complex description of the Dirac theory, electrodynamics, and ultimately the electroweak sector of the standard model. We find that coupling the constructed Dirac spinors to electromagnetism yields the correct magnetic moment for charged spin-1/2 particles. Extending electrodynamics to electroweak theory necessitates an algebraic distinction between the structures of the leptonic and Higgs fields not present in the standard model. The conditions of spontaneous symmetry breaking are explored using an alternative representation of weak isospin and hypercharge equivalent to an irreducible representation of $\mathfrak{su}(2)\oplus\mathfrak{u}(1)$ on $\mathbb{C}^4$. This alternative representation disagrees with the standard model on the overall signs of weak neutral currents. 

\end{abstract}

\maketitle

\section{Introduction}

In the mid 19th century, William Rowan 
Hamilton famously discovered the first 
hyper-complex algebra, which is known as 
the quaternions, $\mathbb{H}$. Only a few decades later, William Kingdon 
Clifford generalized the quaternions 
and all hyper-complex algebras in 
formulating what he called geometric 
algebra – which we now refer to 
as Clifford algebra. Since these 
discoveries, the many translations 
between structures in theoretical 
physics and hyper-complex algebras 
have drawn interest and been studied 
at great length. Paul Dirac 
himself wrote a paper on the use of 
quaternions to describe Lorentz 
transformations \cite{Dirac1944}. For readers already 
familiar with the symmetry groups 
encountered in particle physics,  the first impression 
is that the quaternions offer a miraculously 
simple new way to communicate 
descriptions of interactions at the 
fundamental level. However, one 
encounters many unpleasantries upon 
attempting to establish in full this 
analogy between the standard, 
spinorial ($\mathbb{C}^2/\mathbb{C}^4$) 
description of fundamental 
interactions and a quaternionic ($\mathbb{H}/\mathbb{H}^2)$ 
description. One such unpleasantry 
is encountered almost immediately at 
the level of first quantization, 
which is that one is often led to 
defining operators that trivially 
commute with all others to conserve 
the probability of a quaternionic 
wave function \cite{Rotelli1989}. 

The possibilities of reformulating 
quantum mechanics and quantum field 
theories on $\mathbb{H}$ alone were 
thoroughly treated by De Leo and 
Rotelli in \cite{Rotelli1989, DeLeoRotelli1995,DeLeoRotelli1996,DeLeo2001,DeLeo1997}, as well as by Morita \cite{Morita1986} and others. 

An alternative hyper-complex ring to 
attempt to map the structures of 
quantum theory onto is the 
complexified quaternions 
$\mathbb{H}\otimes_\mathbb{R}\mathbb{C}$. The first 
impression one receives from the complex-
quaternions may differ from that suggested above 
regarding $\mathbb{H}$ itself. Being twice the real 
dimension of $\mathbb{H}$, this alternative 
offers more structure but comes with its own 
drawbacks. One being that 
$\mathbb{H}\otimes_\mathbb{R}\mathbb{C}$ has zero divisors. Contrary to the experience often had 
working over $\mathbb{H}$ alone and despite 
normalization and other concerns, one finds 
$\mathbb{H}\otimes_\mathbb{R}\mathbb{C}$ to be a 
rather hospitable space for reformulating not only 
quantum mechanics, but a large portion of the 
standard model as well. This approach has been 
embraced in recent decades by Morita 
\cite{Morita2007}, Furey \cite{Furey2015}, and 
others. The main purpose of this work is to 
contribute a novel, complete framework to the 
discussion on hyper-complex 
particle physics as well as to further demonstrate 
the adequacy and elegance of hyper-complex algebras 
in unifying the particles and symmetries of 
fundamental interactions.

We motivate our approach 
with an observation about the favored 
representations of Lie and Clifford 
algebras in quantum mechanics and quantum field 
theory. The typical representation of 
$\mathfrak{su}(2)$ that physicists prefer to use, 
known as the fundamental representation (FR) or Pauli 
matrices, is not quite $\mathfrak{su}(2)$. In fact, 
the FR is more precisely described as $i\mathfrak{su}
(2)$ than $\mathfrak{su}(2)$. The basis in the FR 
square to +1 (as opposed to -1 for the genuine case) 
– computing the Killing form makes this clear. 
Physicists prefer this because in the FR, 
the matrices that span the Lie algebra are both 
unitary and Hermitian – two properties of operators 
that are paramount in quantum mechanics. For this 
reason, when confronted with the task of modeling the 
internal angular momentum (spin) of electrons (and 
later other fermions), the physicists of the early 
20th century were naturally led to define the FR.  
Wolfgang Pauli introduced this representation to 
formalize electron spin in the non-relativistic 
context of $\mathbb{R}^3$ in response to the 
observations made in the Stern-Gerlach experiment. 
This is why we sometimes refer to the FR as the Pauli 
matrices. Not long afterward, Paul Dirac discovered 
the recipe for a theory of the electron consistent 
with relativity, which naturally incorporated the 
matrices introduced by Pauli. Soon after Dirac 
published his theory, a mathematician named Hermann 
Weyl claimed to have found a Lorentz invariant theory 
of spinors with half the structure of Dirac’s – which 
he did – however it was then pointed out by Pauli 
that Weyl’s theory did not account for parity 
symmetry. These discoveries led to three (and 
eventually more) types of spinors to be 
widely known in the physics community. We can 
classify these distinct spinors by the Clifford 
algebras to which they correspond via spin 
representation. The space of Pauli 
spinors corresponds to  the Clifford algebra 
generated by Euclidean $\mathbb{R}^3$, $\text{Cl}_3$. The 
space of Dirac spinors corresponds to the spacetime 
algebra generated by Minkowski spacetime 
$\mathbb{R}^{1,3}$, $\text{Cl}_{1,3}$. 
The space of Weyl spinors corresponds to a projection 
of the Dirac representation onto a definite-chirality 
subspace. In our first task of reformulating these 
essential spinor spaces over the complex-quaternions, 
we find (what is also realized in previous works by 
Morita, Furey, and others) that theories of Pauli and 
Weyl spinors, as well as Maxwell’s equations, can 
live comfortably inside a single copy of 
$\mathbb{H}\otimes_\mathbb{R}\mathbb{C}$, while Dirac 
spinors are more easily translated to restricted 
elements of 
$\mathbb{H}^2\otimes_\mathbb{R}\mathbb{C}$.

\subsection{Remarks on Notation}
The pure quaternions are denoted by 

$\mathbb{H}_p=\{h,j,k \colon h^2=j^2=k^2=hjk=-1\}$

Latin indices $\ell,m,n=1,2,3$ are used exclusively to 
index the pure quaternions $\mathbb{H}_p$, as well as 
similar sets like the pure complex quaternions 
$i\mathbb{H}_p= \{ih,ij,ik\}$ and the three 
spatial dimensions of Minkowski spacetime.

Spacetime indices are Greek, and we take the metric 
$\eta^{\mu\nu}$ with signature $(+1,-1,-1,-1)$.

$q^\star$ – the quaternion conjugate of $q\in\mathbb{H} \space$ or $\space \mathbb{H}\otimes_\mathbb{R}\mathbb{C}$

$\tilde{a}$ – the complex conjugate $a\in\mathbb{C} \space$ or $\space \mathbb{H}\otimes_\mathbb{R}\mathbb{C}$

$v_i^{\star T}$ – the quaternion conjugate transpose of a vector 
$v_i\in\mathbb{H}^n$ or $\mathbb{H}^n\otimes_\mathbb{R}\mathbb{C}$

$\tilde{v}_i^T$ – the complex conjugate transpose of a vector $v_i\in\mathbb{C}^n$ or $v_i\in\mathbb{H}^n\otimes_\mathbb{R}\mathbb{C}$

$b^*$ – the simultaneous complex 
and quaternionic conjugate (double 
or Hermitian conjugate) of 
$b\in\mathbb{H}\otimes_\mathbb{R}\mathbb{C}$

$b_i^\dagger$ – the Hermitian conjugate transpose of $b_i\in\mathbb{H}\otimes_\mathbb{R}\mathbb{C}$

Throughout this work, the quaternions are denoted $\omega_\ell=\{h,j,k\}$ 
where $\ell=1,2,3$ , with the orientation defined by the 
commutator 
$\varepsilon_{\ell mn} \omega_\ell\omega_m=2\omega_n$. The Latin 
indices $\ell,m,n$ when multiplied together will always 
carry the orientation just specified, even when a 
Levi-Civita symbol is not explicitly written next to 
said indices to contract them. Also, because these 
indices index (anti-)Euclidean $\mathbb{R}^3$, we 
reserve the right to stick them in the superscripts or 
subscripts as is convenient in order to tidy up terms, emphasize contractions, etc. It is perhaps useful to 
note that for pure quaternions of the form $\omega_\ell 
v^\ell=hv_x+jv_y+kv_z$, the Clifford product on 
$\mathbb{R}^{0,3}$ is naturally realized in quaternion 
multiplication:

$(\omega_\ell x_\ell )(\omega_\ell y_\ell )=-x_\ell y^\ell+\omega_\ell \varepsilon_{\ell mn}x^m y^n$.

\section{Pauli and Weyl Spinors}

The real span of the Pauli matrices is
more accurately described as $i\mathfrak{su}(2)$ than 
as a genuine representation of $\mathfrak{su}(2)$. We make this 
distinction because the FR introduces a subtle mathematical caveat.  Specifically, $\mathfrak{su}(2)$ is a real Lie algebra, a real 
vector space with a bracket and real 
structure constants: $f_{\ell mn}=-2\varepsilon_{\ell mn}.$  Multiplication by complex numbers would map vectors outside of this algebra. The FR, on the other hand, is a real vector space with a bracket whose structure constants are imaginary: $f_{\ell mn}=2i\varepsilon_{\ell mn}$. Upon establishing the following isomorphisms, our aim is to mimic this structure on the space of complex quaternions.

The real Lie algebra $\mathfrak{su}(2)$ is isomorphic (as a Lie algebra) 
to the vector space of pure quaternions endowed with 
the standard bracket operation, which we denote by $\mathbb{H}_p$. In the same way, $i\mathfrak{su}(2) \cong i\mathbb{H}_p$ is a real vector space isomorphism. Additionally, the isomorphisms $\mathfrak{su}(2) \cong \mathbb{H}_p$ and $i\mathfrak{su}(2) \cong i\mathbb{H}_p$ both preserve the Killing form. We highlight this because the isomorphism often used in the physics literature, $\mathfrak{su}(2) \cong i\mathfrak{su}(2)$ is more superficial, preserving only vector space structure but not the Killing form.
Explicitly,

$\mathfrak{su}(2)=\mathbb{R}\left\{\begin{pmatrix}
0 & i \\
i & 0 
\end{pmatrix} ,  \begin{pmatrix}
0 & -1 \\
1 & 0
\end{pmatrix} , \begin{pmatrix}
i & 0 \\
0 & -i
\end{pmatrix}\right\} \cong \mathbb{R} \{ h,j,k \} = \mathbb{H}_p $

$i\mathfrak{su}(2)=\mathbb{R} \{ \begin{pmatrix}
0 & 1 \\
1 & 0 
\end{pmatrix} ,  \begin{pmatrix}
0 & -i \\
i & 0
\end{pmatrix} , \begin{pmatrix}
1 & 0 \\
0 & -1
\end{pmatrix} \} \cong \mathbb{R} \{ ih,ij,ik \} = i\mathbb{H}_p .$

Consider the elements 
of $\mathfrak{su}(2)$ and $i\mathfrak{su}(2)$ to be 
the images of the standard basis of $\mathbb{R}^3$ under a homomorphism from the Clifford 
algebra $\text{Cl}_3$ to the complex quaternions.  The two bases 
shown above correspond to opposite choices of metric: anti-Euclidean and Euclidean respectively.  In particular,
$\mathfrak{su}(2) \cong \mathbb{H}_p$ and 
$i\mathfrak{su}(2) \cong i\mathbb{H}_p$ are isometric (vector space) isomorphisms between images of $\mathbb{R}^3$ embedded in these algebraic structures.  This is demonstrated in the preservation of the metric signature, as $\mathfrak{su}(2) \cong 
\mathbb{H}_p \cong \mathbb{R}^{0,3}$ and $i\mathfrak{su}(2) 
\cong i\mathbb{H}_p \cong \mathbb{R}^{3,0}$. We now define 
notation suited to our purposes.

Let $\omega_\ell\equiv\{\omega_x=h,\omega_y=j,\omega_z=k\}$ and $\Sigma_\ell\equiv\{\Sigma_x=ih,\Sigma_y=ij,\Sigma_z=ik\}$ denote the standard basis elements for $\mathbb{H}_p$ and $i\mathbb{H}_p,$ respectively.
Since 
$\{\Sigma_m,\Sigma_n\}=2\delta_{mn},$ we identify the components of $\Sigma_\ell$ with the generators of the Clifford algebra $\text{Cl}_{3,0}$ on Euclidean space, $\mathbb{R}^3.$  The resulting algebra isomorphism is $$\text{Cl}_{3,0}=\mathbb{R}\{1,e_1,e_2,e_3,e_1 e_2,e_2 e_3,e_3 e_1,e_1 e_2 e_3 \}$$ $$\cong \mathbb{R}\{1,ih,ij,ik,h,j,k,i\} \cong \mathbb{H}\otimes_\mathbb{R}\mathbb{C}.$$
The even subalgebra $\text{Cl}^+_{3,0}$, being isomorphic to the span of $\{1,h,j,k\},$ contains each $\omega_\ell$.  Under exponentiation, these pure quaternions generate the unit sphere in $\mathbb{H}:$ $$S\mathbb{H}=\{e^{\omega_\ell \theta^\ell}\in\mathbb{H} \space \colon \space \theta^\ell\in\mathbb{R} \text{ for } \ell=1,2,3\} \cong \text{SU}(2) \cong \text{Spin}(3).$$
We embed a vector $x^\ell \in \mathbb{R}^3$ into 
this space via inclusion of $\mathbb{R}^3$ into $\text{Cl}_{3,0}$ followed by the above isomorphism, so that
$x\mapsto\Sigma_\ell x^\ell=x_1 ih+x_2 ij+x_3 ik.$
We call a vector from $\mathbb{R}^3$ embedded in this way a \emph{Pauli vector}.  A Pauli vector’s length squared 
is given by the negative of the quaternionic norm 
$-(\Sigma_\ell x^\ell )^\star (\Sigma_\ell x^\ell ) =x_1^2+x_2^2+x_3^2$. 
This notion of length is invariant under $\text{SO}(3)$ 
transformations, which are carried out in the 
Clifford algebra via conjugation by an element of $S\mathbb{H}.$

The spinor spaces (left ideals) upon which $S\mathbb{H}$ acts consist of full complex quaternions of the form
$\chi=\alpha_0+\alpha_\ell\omega_\ell$ where $\alpha_0,\alpha_\ell \in \mathbb{C}$.  As the coefficients are complex, there is no real inner product, so these Pauli spinors are not real-normalizable. Although complex, the Hermitian inner product on these spinors is preserved under the action of $S\mathbb{H}.$

We define analogs to the Weyl spinor formalism by identifying the 
unit scalar with the time basis vector as 
$$\Sigma^\mu=\{\Sigma^0 = 1 ,\Sigma^\ell=i\omega_\ell \}.$$ 
Embedding an arbitrary four-vector $x^\mu \in 
\mathbb{R}^{1,3}$ as $\Sigma^\mu x_\mu=x_0-i\omega_\ell x_\ell = x_0-ihx_1-ijx_2-ikx_3$, one may verify that the length given by $$(\Sigma^\mu x_\mu )^\star (\Sigma^\mu x_\mu )= x_0^2-x_\ell^2=x_0^2-x_1^2-x_2^2-x_3^2$$ is invariant under $\text{SO}(1,3)$ transformations implemented via conjugation by $\text{Spin}(1,3)$, which here we call the complex-quaternion sphere denoted by
$$S(\mathbb{H}\otimes_R\mathbb{C})= \{ e^{\omega_\ell\beta^\ell} : \beta_\ell \in \mathbb{C} \} \cong \text{SL}(2, \mathbb{C}) \cong \text{Spin}(1,3).$$ Conjugation by an element $e^{\omega_\ell \beta^\ell}\in 
S(\mathbb{H}\otimes_\mathbb{R} \mathbb{C})$ transforms a vector as
$(\Sigma^\mu x_\mu )' = e^{\omega_\ell \beta^\ell}  
(\Sigma^\mu x_\mu ) e^{-\omega_\ell \beta^\ell} 
=x_0+ie^{2\omega_\ell \beta^\ell} \omega_\ell x_\ell$,
and thus preserves the spacetime interval $$((\Sigma^\mu x_\mu)^\star(\Sigma^\mu x_\mu ))'= 
(x_0-i\omega_\ell x_\ell e^{-2\omega_\ell \beta^\ell} )
(x_0+ie^{2\omega_\ell \beta^\ell} \omega_\ell x_\ell )= x_0^2-
x_\ell^2.$$

A representation of the Lorentz group isomorphic but inequivalent to that above is the complex conjugate of $\Sigma^\mu$, 
$\tilde{\Sigma}^\mu=\{\tilde{\Sigma}^0=1, 
\tilde{\Sigma}^\ell = -i\omega_\ell\}$. The spin group generated by 
the conjugate generators, relative to the former rep, contains general elements of the form 
$e^{\omega_\ell 
\tilde{\alpha}^\ell} \in 
S(\mathbb{H}\otimes_\mathbb{R} \tilde{\mathbb{C}})$.
The distinction between the two representations 
$S(\mathbb{H}\otimes_\mathbb{R}\mathbb{C})$ and 
$S(\mathbb{H}\otimes_\mathbb{R} \tilde{\mathbb{C}})$  
is that their respective boost transformations 
(of the form $e^{\pm i\omega_\ell \theta^\ell}$ for 
$\theta^\ell\in\mathbb{R})$ are opposite. We identify the spin group generated by the components of $\Sigma^\mu$ 
as a right-chiral representation of the Lorentz group and 
those of $\tilde{\Sigma}^\mu$ with a left-chiral representation.

\section{The Dirac Theory}

We construct the Dirac gamma matrices by taking the direct sum of the two oppositely chiral representations of the Lorentz group generated by $\Sigma^\mu$ and $\tilde{\Sigma}^\mu$.
One version of this direct sum is analogous to the Weyl (chiral) representation of the gamma matrices:

$$\Gamma^\mu = \begin{pmatrix}
0 & \Sigma^\mu \\
\tilde{\Sigma}^\mu & 0 
\end{pmatrix}
= \left( \begin{pmatrix}
0 & 1 \\
1 & 0 
\end{pmatrix} , 
\begin{pmatrix}
0 & \Sigma^\ell \\
-\Sigma^\ell & 0 
\end{pmatrix} \right)^T.$$

Another, the Dirac representation, is obtained by 
the unitary transformation $U=\frac{1}{\sqrt{2}} 
\begin{pmatrix}
    1 & 1 \\
    -1 & 1
\end{pmatrix}$, and is given by
$\Gamma^\mu = \left( \begin{pmatrix}
1 & 0 \\
0 & -1 
\end{pmatrix} , 
\begin{pmatrix}
0 & \Sigma^\ell \\
-\Sigma^\ell & 0 
\end{pmatrix} \right)^T$.

The Dirac spinors are denoted by $\psi= (\psi_\ell, \psi_R)^T \in \mathbb{H}^2 \otimes_\mathbb{R} \mathbb{C}$. In this notation, the Dirac adjoint takes on its standard form,
$\bar{\psi}=\bar{\psi}^\dagger\Gamma^0$, where the 
Hermitian conjugate (conjugate transpose) is used.

An advantage of $\mathbb{H}\otimes_\mathbb{R} \mathbb{C}=\mathbb{R}\{h,j,k,i\}$ at the level of first quantization is the presence of a globally commuting root of -1.
Thereby we define the four-momentum operator as 
$\hat{p}_\mu=i\partial_\mu,$ which is complex-Hermitian but 
not quaternionic-Hermitian.
The momentum operator must be defined this way in order
to be able to define the Dirac operator in the form
$$(i\Gamma^\mu \partial_\mu-m)\psi=0.$$
As the momentum operator is complex-Hermitian, we guess 
solutions of the form $\psi_+(x)= u(p)e^{-ip_\mu x^\mu}$ 
and $\psi_- (x)= v(p)e^{ip_\mu x^\mu}$ correspond to positive and 
negative energy solutions, respectively. Under the action of the operators $\hat p^\ell=-i\partial^\ell$ and 
$\hat{E}=i\partial_t$, the eigenvalues of $\psi_+$ and 
$\psi_-$ are $p_\ell,E$ and $-p_\ell,-E$ respectively. 
The coefficients $u(p)$ and $v(p)$ of each ansatz 
are functions of their respective 
solution’s momentum which are valued in $\mathbb{H}^2 \otimes_\mathbb{R} \mathbb{C}$.  We assert that the solutions will be compatible with the Hermitian norm mentioned above
$\bar{\psi}\psi=\psi_\ell^*\psi_\ell-\psi_R^*\psi_R$.

The Dirac equation yields two positive and two negative energy solutions
\begin{align*}
\psi_1&= \begin{pmatrix}
    1 \\
    \frac{-i\omega_\ell p_\ell}{E+m}
\end{pmatrix} e^{-ip_\mu x^\mu}, \quad \psi_2= \begin{pmatrix}
    \frac{-i\omega_\ell p_\ell}{E-m} \\
    1
\end{pmatrix} e^{-ip_\mu x^\mu}, \\
\psi_3&= \begin{pmatrix}
        \frac{-i\omega_\ell p_\ell}{E+m} \\
    1
\end{pmatrix} e^{ip_\mu x^\mu},\quad\psi_4= \begin{pmatrix}
    1 \\
    \frac{-i\omega_\ell p_\ell}{E-m}
\end{pmatrix} e^{ip_\mu x^\mu}.
\end{align*}
We apply the Feynman-St\"{u}ckelberg interpretation, mapping negative-energy solutions to anti-particles, by sending $E\rightarrow-E$ and $p\rightarrow-p.$ We see that this transformation sends $\psi_2\rightarrow\psi_3$ and $\psi_4\rightarrow\psi_1$. Thus, we have two independent solutions, one matter and one antimatter, given by 
$$\chi= \begin{pmatrix}
    1 \\
    \frac{-i\omega_\ell p_\ell}{E+m}
\end{pmatrix} e^{-ip_\mu x^\mu} \space , \space \phi= \begin{pmatrix}
    \frac{-i\omega_\ell p_\ell}{E+m} \\
    1
\end{pmatrix} e^{ip_\mu x^\mu}.$$ 
We can turn these spinor solutions into corresponding solutions in the Weyl representation via the inverse of the aforementioned unitary transformation, i.e.
$\chi_D \rightarrow \chi_W= U^\dagger\chi_D = \frac{1}{\sqrt{2}} \begin{pmatrix}
    1 & -1 \\
    1 & 1
\end{pmatrix} \begin{pmatrix}
    1 \\
    \frac{-i\omega_\ell p_\ell}{E+m}
\end{pmatrix}e^{-ip_\mu x^\mu} = \frac{1}{\sqrt{2}} \begin{pmatrix}
    1 + \frac{i\omega_\ell p_\ell}{E+m} \\
    1 - \frac{i\omega_\ell p_\ell}{E+m}
\end{pmatrix}e^{-ip_\mu x^\mu} $.

Returning to the Dirac representation, the charge conjugation operator is given by $C= \begin{pmatrix}
    0 & -1 \\
    1 & 0
\end{pmatrix}$, and one may verify that $\chi\rightarrow C\Gamma^0 \chi^* = \phi$. The time-reversal operator is the complex-Hermitian matrix $T = \begin{pmatrix}
    0 & -i \\
    i & 0
\end{pmatrix}$.  One may verify that with these representations of C, P, and T the Dirac equation is CPT invariant.

Using the chirality operator $\Gamma^5=i\Gamma^0 \Gamma^1 \Gamma^2 \Gamma^3= \begin{pmatrix}
    -1 & 0 \\
    0 & 1
\end{pmatrix}$ we project the spinor onto its left- and right-chiral components $\chi_\ell=\frac{(1-\Gamma^5)}{2} \chi$ and $\chi_R=\frac{(1+\Gamma^5)}{2} \chi$, and $\Gamma^5 \chi_\ell=-\chi_\ell$ and $\Gamma^5 \chi_R=+\chi_R$. 

\section{Spin, Helicity, and Magnetic Moment}
The Schr\"{o}dinger form of the Dirac equation is $i\partial_t \psi=(\alpha^\ell p_\ell+\beta m)\psi$, where    $\alpha^\ell=\Gamma^0 \Gamma^\ell= \begin{pmatrix}
    0 & i\omega_\ell \\
    i\omega_\ell&0
\end{pmatrix}$ and $\beta=\Gamma^0$, and the Hamiltonian is
$$H=\alpha^\ell p_\ell + \beta m = \begin{pmatrix}
    m & i\omega_\ell p_\ell \\
    i\omega_\ell p_\ell & -m
\end{pmatrix}.$$
Define the spin operator as $\vec{S}=S_\ell= \frac{1}{2} \begin{pmatrix}
    \Sigma_\ell & 0 \\
    0 & \Sigma_\ell
\end{pmatrix} = \frac{1}{2}\begin{pmatrix}
    i\omega_\ell & 0 \\
    0 & i\omega_\ell
\end{pmatrix}$. The total angular momentum given by $\vec{J}=\vec{L} + \vec{S}$ where $\vec{L} = L_\ell =\varepsilon_{\ell mn}r^m  p^n$, which commutes with the Hamiltonian since $[H,L_\ell]=[\alpha^k p_k, L_\ell]=i\varepsilon_{\ell mn}\alpha^m p^n$ and $[H, S_\ell]=-i\varepsilon_{\ell mn}\alpha^m p^n$.

The helicity operator is given by $h=\frac{S_\ell p^\ell}{\sqrt{p_\ell p^\ell}} = \frac{1}{2\sqrt{p_\ell p^\ell}} \begin{pmatrix}
    i\omega_\ell p_\ell & 0 \\
    0 & i\omega_\ell p_\ell
\end{pmatrix}$, which commutes with the Hamiltonian.

\subsection{Spin Eigenstates}

The previously identified matter and antimatter 
solutions to the Dirac equation, $\chi$ and $\phi$, are 
not eigenstates of spin/helicity. Restricting motion to be in the z-direction reduces the helicity operator to the z-component of spin $S_z= \frac12\begin{pmatrix}
    ik & 0 \\
    0 & ik
\end{pmatrix}.$  However the spinors
$\chi= \begin{pmatrix}
    1 \\
    \frac{-ikp_z}{E+m}
\end{pmatrix} , \phi = \begin{pmatrix}
    \frac{-ikp_z}{E+m} \\
    1
\end{pmatrix},$ are still not eigenstates. To resolve this, we employ certain idempotents from $\mathbb{H} \otimes_\mathbb{R}\mathbb{C}$ as coefficients. These coefficients, found independently here, are of the same type as those found by Furey in \cite{Furey2015}. Notice that restricting spinor solutions to travel in the z-direction happens to send the spinor components to $\mathbb{C}\{1,k\}$, which is an Abelian subalgebra of $\mathbb{H} \otimes_\mathbb{R}\mathbb{C}$. This subalgebra is also four-dimensional over $\mathbb{R}$, corresponding to the number of linearly independent helicity states available to a matter particle and its antimatter counterpart. We choose orthogonal idempotents from this subalgebra that commute with the spinors.

Define the following matter and antimatter spin eigenstates
\begin{align*}
    \chi_\uparrow&=\frac{1}{2}(1+ik)\chi,  \quad\chi_\downarrow=\frac{1}{2}(i+k)\chi,\\ \phi_\uparrow&=\frac{1}{2}(1-ik)\phi, \quad
\phi_\downarrow=\frac{1}{2}(-i+k)\phi.
\end{align*}
We chose these idempotents so that the eigenvalues of $S_z$ corresponding to $\chi_\uparrow,$ $\chi_\downarrow,$ $\phi_\uparrow,$ and $\phi_\downarrow$ are $+1/2$, $-1/2$, $-1/2$, and $+1/2$ respectively.  It follows that the pairs $(\chi_\uparrow, \chi_\downarrow),$ $(\phi_\uparrow, \phi_\downarrow)$ are orthogonal. We define a real norm on the Dirac spinors as the real part of the Hermitian inner product $Re(\bar\chi \chi)=Re(\chi^\dagger \Gamma^0 \chi).$

\subsection{Magnetic Moment}

To investigate the magnetic moment of our spinors, we minimally couple a particle with charge $e$ to an electromagnetic field $A_\mu=(\varphi,A_\ell)$ yielding $\partial_\mu \rightarrow \partial_\mu+ieA_\mu$ corresponding to a kinematic momentum $\pi_\mu=p_\mu-eA_\mu$. The coupled Dirac equation
$(i\Gamma^\mu (\partial_\mu+ieA_\mu)-m)\psi=(\Gamma^\mu (p_\mu-eA_\mu)-m)u(p)=0$ decomposes into two coupled equations, one for each spinor component:
\begin{align*}
    (E-e\varphi-m) \alpha&=i\omega_\ell\pi^\ell\beta \space \\
(E-e\varphi+m)\beta&=i\omega_\ell\pi^\ell\alpha.
\end{align*}
Defining the relativistic kinetic energy as $T=E-e\varphi-m$ and performing elimination gives 
$$(i\omega_\ell\pi^\ell)^2\alpha=(E-e\varphi+m)T\alpha$$
Using $(\omega_\ell x^\ell )(\omega_\ell y^\ell )=-x_\ell y^\ell+\omega_\ell \varepsilon^\ell_{mn}x^m y^n$ and recognizing $-\frac{i}e\varepsilon^\ell_{mn}\pi^m\pi^n=-\frac i{2e}\varepsilon^\ell_{mn}[\pi^m,\pi^n]=B^\ell$ as the magnetic field, we have $$[(p_\ell-eA_\ell)^2-e\Sigma_\ell B^\ell ]\alpha=(E-e\varphi+m)T\alpha.$$

In the non-relativistic limit, i.e., $T<<m$ the kinetic energy becomes $T\approx\frac{1}{2m}[(p_\ell-eA_\ell)^2-e\Sigma_\ell B^\ell]$. Thus, the total energy is $H=T+e\varphi$ and we get the Pauli equation for the spinor $\alpha$ 
$$\left[\frac{(p_\ell-eA_\ell)^2}{2m}-\frac{e}{2m}\Sigma_\ell B^\ell +e\varphi\right]\alpha=E\alpha$$
The magnetic moment of this complex-quaternionic spin-$1/2$ particle is 

$\vec{\mu}=\mu_\ell=\frac{e}{2m}\Sigma_\ell$, and its energy is $-\vec{\mu} \cdot \vec{B}=-\mu_\ell B^\ell$.

\section{Maxwell's Equations and Electrodynamics}
In full, the spin representation of the spacetime algebra we have been considering is spanned by the elements:
$$\mathbb{R}\{I,\Gamma^\mu,\Gamma^\mu \Gamma^\nu,\Gamma^\mu \Gamma^\nu \Gamma^\lambda,\Gamma^4\},$$
where $\Gamma^4=\Gamma^0\Gamma^1\Gamma^2\Gamma^3$ is the pseudoscalar of the Clifford algebra.
The inner product is the anticommutator and the wedge product is the commutator. The Clifford product is given by
$\Gamma^\mu \Gamma^\nu=\frac12\{\Gamma^\mu,\Gamma^\nu\}+\frac12[\Gamma^\mu,\Gamma^\nu]=\eta^{\mu\nu}+\Gamma^\mu\wedge\Gamma^\nu$.
We embed a four-vector potential as $A^\mu=(\varphi,\vec{A}) \rightarrow A=\Gamma^\mu A_\mu= \begin{pmatrix}
    \varphi & -i\omega_\ell A_\ell \\
    i\omega_\ell A_\ell & -\varphi
\end{pmatrix}$. 
Using the embedded spacetime derivative $d=\Gamma^\mu\partial_\mu,$ the field strength is given by $F=dA= \begin{pmatrix}
    -\omega_\ell B_\ell & i\omega_\ell E_\ell \\
    i\omega_\ell E_\ell & -\omega_\ell B_\ell
\end{pmatrix} = \Gamma^0\Gamma^\ell E_\ell+\Gamma^m\Gamma^nB_\ell.$  Electric and magnetic fields are recognized as $E_\ell=-\partial_0 A_\ell-\partial_\ell\varphi$ and $B_\ell=\varepsilon_{\ell mn}\partial^m A^n$, and we have chosen the Lorenz gauge $\partial_\mu A^\mu=\partial_0\varphi+\partial_\ell A^\ell=0.$
The source-free Maxwell equations are given by $dF=0$. To add sources, we embed a four-current as $j=\Gamma^\mu j_\mu=\Gamma^0 \rho+\Gamma^\ell j_\ell= \begin{pmatrix}
    \rho & -i\omega_\ell j_\ell \\
    i\omega_\ell j_\ell & -\rho
\end{pmatrix}$. Maxwell's equations with source are $$dF=j \space .$$ 
Written out in full they read:
 $$\begin{pmatrix}
    \partial_\ell E_\ell-\omega_\ell (\varepsilon_{\ell mn}\partial^m E^n+\partial_0 B_\ell) & i\partial_\ell B_\ell+i\omega_\ell (\partial_0 E_\ell -\varepsilon_{\ell mn}\partial^m B^n ) \\
    i\omega_\ell (\varepsilon_{\ell mn}\partial^m B^n -\partial_0 E_\ell )-i\partial_\ell B_\ell & \omega_\ell (\varepsilon_{\ell mn}\partial^m E^n+\partial_0 B_\ell)-\partial_\ell E_\ell
\end{pmatrix}= \begin{pmatrix}
    \rho & -i\omega_\ell j_\ell \\
    i\omega_\ell j_\ell & -\rho
\end{pmatrix}.$$
Equating the scalar coefficients on each side yields Gauss's laws, while Faraday's and Ampere's laws follow from equating the coefficients of $\omega_\ell.$
The square of the field strength is $F^2=\begin{pmatrix}
    -B_\ell^2+E_\ell^2 & 0 \\
    0 & -B_\ell^2+E_\ell^2
\end{pmatrix}$.

\newcommand{\Lag}{\mathscr{L}}

Using the embedded derivative, the free Dirac Lagrangian
density is 
$\Lag=i\bar{\psi}d\psi-m\bar{\psi}\psi$. Defining the 
gauge covariant derivative as 
$D=\Gamma^\mu D_\mu=d+iqA$, electrodynamics takes the 
form
$$\Lag=i\bar{\psi} D\psi-m\bar{\psi}\psi-\frac{1}{4}TrF^2.$$

We check charge conservation by varying with respect to an infinitesimal gauge transformation $\delta\lambda$ so that
\begin{align*}\delta\Lag &=
(\delta\lambda) \bar{\psi} \Gamma^\mu D_\mu \psi -\bar{\psi}\Gamma^\mu D_\mu (\delta\lambda\psi)\\ &= -
\bar{\psi} \Gamma^\mu \psi (\partial_\mu \delta\lambda)\\& = \delta\lambda\partial_\mu (\bar{\psi}\Gamma^\mu \psi).
\end{align*}
Setting $\delta\Lag=0$ yields $\partial_\mu (\bar{\psi}\Gamma^\mu \psi)=0.$
Thus, the conserved current associated with the $\text{U}(1)$ gauge symmetry is 
$$J=j^\mu=\bar{\psi}\Gamma^\mu\psi.$$

\section{Electroweak Theory}
\subsection{The Standard Choice}
The constructed complex quaternion Dirac spinor formalism is compatible with the standard representation of weak isospin and hypercharge because it is a subalgebra of $\mathfrak{gl}_2(\mathbb{H})\otimes_\mathbb{R}\mathbb{C}$ and naturally transforms complex quaternionic doublets.

$$\mathfrak{su}(2)_L= \left\{ \sigma_1=\frac{1}{2}\begin{pmatrix}
    0 & 1 \\
    1 & 0
\end{pmatrix} , \sigma_2=\frac{1}{2}\begin{pmatrix}
    0 & i \\
    -i & 0
\end{pmatrix}, \sigma_3=\frac{1}{2}\begin{pmatrix}
    1 & 0 \\
    0 & -1
\end{pmatrix} 
\right\}$$  $$\mathfrak{u}(1)_Y=\left\{ Y=\frac{1}{2}\begin{pmatrix}
     1 & 0 \\
     0 & 1
 \end{pmatrix} \right\}$$

Using the standard choice, the conditions of spontaneous symmetry breaking yield currents and coupling terms that are in agreement with the Standard Model.

\subsection{An Alternative Choice}
As seen in previous sections, the standard representations of weak isospin and hypercharge are not the only representations of $\mathfrak{su}(2)$ and $\mathfrak{u}(1)$ in the 32-dimensional space of matrices of $\mathfrak{gl}_2\mathbb{H}\otimes_\mathbb{R}\mathbb{C}$. However, there is only one other representation of these algebras that, when combined in a gauge covariant derivative, yield results similar to the FR upon imposing conditions of spontaneous symmetry breaking (SSB). Among these similar results are the currents containing mixtures of left-chiral leptons in two charged interactions and non-mixing in two neutral interactions.

The representation of weak isospin is structurally distinct from all other 
isomorphic subalgebras in 
$\mathfrak{gl}_2(\mathbb{H})\otimes_\mathbb{R} \mathbb{C}$ encountered in this work.  All such subalgebras are equivalent to the irreducible representation of $\mathfrak{su}(2)$ on $\mathbb{C}^2$ or are reducible. The following representation of $\mathfrak{su}(2)$ is unique in its equivalence to an irreducible representation of $\mathfrak{su}(2)$ on $\mathbb{C}^4$. In the fundamental representation, these matrices are

$$\mathfrak{su}(2)= \left\{ x_1=\frac12\begin{pmatrix}
    0 & -h \\
    h & 0
\end{pmatrix} , x_2=\frac12\begin{pmatrix}
    0 & -j \\
    j & 0
\end{pmatrix}, x_3=\frac12\begin{pmatrix}
    ik & 0 \\
    0 & ik
\end{pmatrix} 
 \right\}$$
and
$$\mathfrak{u}(1)=\left\{ y=\frac12\begin{pmatrix}
     ik & 0 \\
     0 & -ik
 \end{pmatrix} \right\}$$
The Lie group generated by these algebras naturally acts on any vector subspace of $\mathbb{H}^2\otimes_\mathbb{R} \mathbb{C}$.

The electroweak Lagrangian prior to spontaneous symmetry 
breaking (SSB) is invariant under the gauge group $\text{SU}(2) 
\otimes U(1),$ the total electroweak Lagrangian is conventionally written as:

$$\Lag\indices{_E_\ell_e_c_t_r_o_w_e_a_k} = \Lag\indices{_F_e_r_m_i_o_n}+\Lag\indices{_G_a_u_g_e}+\Lag\indices{_H_i_g_g_s}+\Lag\indices{_Y_u_k_a_w_a}$$

We restrict our discussion to the first 
generation of leptons. The left-chiral electron and neutrino form a weak isospin doublet $ L = \begin{pmatrix} 
\nu_L \\
e_L
\end{pmatrix} $. This doublet is an eigenvector of the third component of weak isospin $x_3=\frac{1}{2}\begin{pmatrix}
    ik & 0 \\
    0 & ik
\end{pmatrix}$ with eigenvalue $\frac{ik}{2}.$  As $x_3$ is a multiple of the identity, both components of the isospin doublet have the same eigenvalue. This differs from the conventional formulation, in which the diagonal weak isospin generator is $\frac{1}{2}\sigma_3$ and assigns opposite sign eigenvalues $(\pm\frac{1}{2})$ to the two components.  As shown below, after symmetry breaking this difference leads to a disagreement with the standard model on the signs of the currents coupled to the Z boson.

The right-chiral electron $e_R$ transforms as a weak hypercharge 
singlet.  The fermion sector consists of parity-
violating gauge covariant kinetic terms given by 
\begin{align*}
    \Lag{_F^L} &= i L{^\dagger} \Sigma{^\mu} (\partial{_\mu} + \frac{1}{2} g x{_\ell} X{^\ell_\mu} - \frac{1}{2} g' y B{_\mu} ) L,\\ 
 \Lag{_F^R} &= i e{_R^\dagger} \Sigma{^\mu} (\partial{_\mu} - g' y_{22}B{_\mu} ) e_R.
\end{align*}
Note that $i$ does not multiply the vector potentials in either gauge covariant derivative. This choice is intentional. The gauge covariant derivative must be defined this way to admit a consistent definition of electric charge after symmetry breaking.

The Higgs sector is given by
$$\Lag_{Higgs}=(D_\mu \phi)^\dagger D^\mu \phi+m^2\phi^\dagger \phi -\lambda (\phi^\dagger \phi)^2,$$ 
where $D_\mu \phi=(\partial{_\mu} + \frac{1}{2} ig x{_\ell} X{^\ell_\mu} + \frac{1}{2} ig' y B{_\mu} )\phi.$
To implement spontaneous symmetry breaking of 
three of the four $\text{SU}(2)$ generators, the Higgs 
field before SSB must have four real degrees of freedom, leaving a single scalar field after SSB. The only vector 
subspace of $\mathbb{H}^2\otimes_\mathbb{R} \mathbb{C}$ that transforms non-trivially under this $2\times2$ representation and has at most four real dimensions is 
$\mathbb{C}^2$. This choice algebraically distinguishes the Higgs field from the chiral components of the fermion 
fields. This distinction is present regardless of the chosen representation of $\mathfrak{su}(2)\oplus\mathfrak{u}(1)$ as long as it is a subalgebra of $\mathfrak{gl}_2\mathbb{H}\otimes_\mathbb{R} \mathbb{C}$. In the standard formulation of electroweak theory, both the Higgs and the chiral components of fermions take values in $\mathbb{C}^2$.  However, the structure of the Higgs is unique in this framework. The broader physical implications of this distinction remain an open question.

Finally, the Yukawa coupling is
$\Lag_{Yukawa}=-G_e(e_R^* \phi^\dagger L+L^\dagger \phi e_R)$.

\subsection{The Alternative Choice after SSB}
Expanding the left-chiral fermion sector gives three terms, the kinetic terms, the charged currents, and the neutral currents: $\Lag^L_F=K_L+C_L+N_L$.
$$K_L=i\nu_L^*\tilde{\Sigma}^\mu\partial_\mu\nu_L+ie_L^*\tilde{\Sigma}^\mu\partial_\mu e_L$$
$$C_L=\frac{1}{2}ig\left[\nu_L^*\tilde{\Sigma}^\mu(-hX^1_\mu -jX^2_\mu)e_L+e_L^* \tilde{\Sigma}^\mu(hX^1_\mu+jX^2_\mu)\nu_L \right]$$
$$N_L=-\frac{1}{2}\nu_L^* \tilde{\Sigma}^\mu(kgX_\mu^3-kg'B_\mu)\nu_L-\frac{1}{2}e_L^* \tilde{\Sigma}^\mu(kgX_\mu^3+kg'B_\mu)e_L$$

Using the definitions $W_\mu^\pm=\mp i\frac{(hX_\mu^1+jX_\mu^2)}{\sqrt{2}}$,  $X^3_\mu=\sin\theta A_\mu+\cos\theta Z_\mu$, and $B_\mu=\cos\theta A_\mu-\sin\theta Z_\mu$ where $\theta$ is the Weinberg angle defined by $(\tan\theta=\frac{g'}{g})$, the charged and neutral currents become
$$C_L=\frac{g}{\sqrt{2}}\left[\nu_L^*\tilde{\Sigma}^\mu W^+_\mu e_L+e_L^* \tilde{\Sigma}^\mu W^-_\mu\nu_L \right]$$
$$N_L=-\frac{1}{2}\nu^*_L\tilde{\Sigma}^\mu k\left[g(\sin\theta A_\mu+\cos\theta Z_\mu) -g'(\cos\theta A_\mu -\sin\theta Z_\mu)\right]\nu_L$$ $$-\frac{1}{2}e_L^*\tilde{\Sigma}^\mu k\left[g(\sin\theta A_\mu+\cos\theta Z_\mu) +g'(\cos\theta A_\mu -\sin\theta Z_\mu)  \right]e_L.$$
Expanding right-chiral fermion term yields 
$$\Lag^R_F=K_R+N_R= ie_R^*\Sigma^\mu \partial_\mu e_R-g'e_R^*\Sigma^\mu k(\sin\theta A_\mu+\cos\theta Z_\mu)e_R.$$
Adding together the right- and left-chiral fermion Lagrangians and using the chiral projection identities $f_L^*\tilde{\Sigma}^\mu f_L=\bar{\psi}_f\Gamma^\mu\frac{1-\Gamma^5}{2}\psi_f$ and $f_R^*\Sigma^\mu f_R=\bar{\psi}_f\Gamma^\mu\frac{1+\Gamma^5}{2}\psi_f$ for $f=$ $e$ or $\nu$, we can write the sum of kinetic terms and sum of neutral currents as
$$K_L+K_R=i\nu_L^*\tilde{\Sigma}^\mu\partial_\mu\nu_L+i\bar{\psi}_e\Gamma^\mu\partial_\mu\psi_e,$$
$$N_L+N_R=-\frac{g}{2}(\sin\theta A_\mu +\cos\theta Z_\mu) \left[\bar{\psi}_\nu\Gamma^\mu k\frac{1-\Gamma^5}{2}\psi_\nu+\bar{\psi}_e\Gamma^\mu k\frac{1-\Gamma^5}{2}\psi_e\right]$$
$$+g'(\cos\theta A_\mu-\sin\theta Z_\mu)\left[\frac{1}{2}(\bar{\psi}_\nu\Gamma^\mu k\frac{1-\Gamma^5}{2}\psi_\nu+\bar{\psi}_e\Gamma^\mu k\frac{1-\Gamma^5}{2}\psi_e)-\bar{\psi}_e\Gamma^\mu\psi_e  \right]$$
The definition of electric charge is $e=g'\cos\theta=g\sin\theta$. The electromagnetic source current that we derived variationally in the previous section (with appropriate adjustments to the U(1) Lie algebra) is present: $J^{EM}_\mu=-\bar{\psi}_e\Gamma^\mu k\psi_e=\sum\limits_{f} q_f\bar{\psi}_f\Gamma^\mu k\psi_f$ where $q_f=-1,0$ for $f=e,\nu$ respectively . The weak neutral current we see is $J_\mu^3=\sum_f I_f^3\bar{\psi}_f\Gamma^\mu k\frac{1-\Gamma^5}{2}\psi_f$ where $I_f^3=\frac{1}{2}$ for both $f=e,\nu$.

With these definitions the neutral currents can be written as
$$N_L=eA^\mu J^{EM}_\mu -\frac{g}{\cos\theta}Z^\mu \left[ J_\mu^3-\sin^2\theta J^{EM}_\mu\right].$$
As we anticipated earlier when considering the eigenvalues of the components of the isospin doublet, the terms where currents couple to the Z boson are a factor of $-1$ off from normal. To rationalize this, we look to the $+$ sign on the photon coupling $eA^\mu J^{EM}_\mu$ and observe that it indicates that the force between particle/antiparticle pairs such as $(e^-, e^+)$ is attractive. The opposite sign on the weak neutral coupling $-\frac{g}{\cos\theta}Z^\mu \left[ J_\mu^3-\sin^2\theta J^{EM}_\mu\right]$ implies that the Z-boson mediates a repulsive force between pairs ($e_L$, $e^*_L$) and ($\nu_L$, $\nu^*_L$), and an attractive force between ($\psi_e$, $\bar\psi_e$).
Another oddity encountered using this representation of $\mathfrak{su}(2)$ in place of weak isospin is that the inner product between the resulting $W^{+\mu}$ and $W^{-\mu}$ bosons is negative definite, $W^+_\mu W^{-\mu}<0$. After symmetry breaking, this representation's analogue of electric charge is defined as $Q=\begin{pmatrix}
    0 & 0 \\
    0 & k
\end{pmatrix}$. If we extend the linearity to allow for simultaneous left and right actions by pure quaternions, linear combinations yield the broken generators corresponding to the $W_\mu^+$ and $W_\mu^-$ are $w^+=\frac{1}{2}\begin{pmatrix}
    0 & -i(h+j) \\
    0 & 0
\end{pmatrix}$ and $\tilde{w}^T=\frac{1}{2}\begin{pmatrix}
    0 & 0 \\
    i(h+j) & 0
\end{pmatrix}$ respectively. Under adjoint action by $Q$, the right eigenvalues of $w$ and $\tilde{w}^T$ are $+k$ and $-k$ respectively. 
The consequences of this alternative representation may be explored further in a future work

The Higgs field, after developing a vacuum expectation value (VEV) and selecting the Weinberg gauge, takes the following form
$$\phi=\begin{pmatrix}
    \phi_1 \\
    \phi_2
\end{pmatrix} \in \mathbb{C}^2 \xrightarrow[]{SSB} \phi'=\frac{1}{\sqrt{2}} \begin{pmatrix}
    0 \\
    H(x)+v
\end{pmatrix}$$
The field $H(x)$ is a real scalar field (the Higgs field post-SSB), and $v=\frac{\pm m}{\sqrt{2\lambda}}$ is the VEV. After carrying this out, expanding the Higgs kinetic term yields the mass couplings of the weak bosons as well as the various terms coupling the Higgs field and the weak bosons.
$$\Lag^{Broken}_{Higgs-Kinetic}=(D_\mu (H+v))^\dagger D^\mu (H+v)$$
$$=\frac{1}{2}\partial_\mu H\partial^\mu H+\frac{1}{8}g^2(H+v)^2\left [ -(X_\mu^1 X_1^\mu+X_\mu^2 X_2^\mu)+(g X_\mu^3-g'B_\mu)^2 \right]$$
$$=\frac{1}{2}\partial_\mu H\partial^\mu H+\frac{1}{4}g^2(H+v)^2 W_\mu^+ W^{-\mu}+\frac{g^2+g'^2}{8}(H+v)^2Z_\mu Z^\mu $$
Substituting in the mass definitions $m_W=\frac{gv}{2}$ and $m_Z=\frac{v\sqrt{g^2+g'^2}}{2}$, the Higgs sector becomes
$$\Lag^{Broken}_{Higgs-Kinetic}=\frac{1}{2}\partial_\mu H\partial^\mu H+m_W^2W_\mu^- W^{+\mu}+\frac{1}{2}m_Z^2Z_\mu Z^\mu+\Lag_{HV}$$ where $\Lag_{HV}=\frac{1}{v^2} \left[ m_W^2W_\mu^+ W^{-\mu}+\frac{m_Z^2}{2}Z_\mu Z^\mu\right](H^2+2Hv).$ 

Taking the usual electron mass $m_e=\frac{G_ev}{\sqrt{2}}$, the broken Yukawa coupling is $$\Lag_{Yukawa}^{Broken}=-\frac{G_e}{\sqrt{2}}(e_R^* (H+v) e_L+e_L^* (H+v) e_R)=-m_e\bar{\psi}_e\psi_e-\frac{gm_e}{2m_W}H\bar{\psi}_e\psi_e$$

\section{Conclusion} 
Through mirroring the subtle difference between the Lie algebra $\mathfrak{su}(2)$ and the version of it that physicists prefer to use (the Pauli matrices), we were able to construct a complex quaternionic Dirac spinor theory compatible with the standard model gauge field interactions.

The table below compares the results found here to the standard results.
\begin{table}[h]
    \centering
    \begin{tabular}{|c|c|c|}\hline
     & Standard & Complex Quaternionic\\ \hline
        Dirac Spinors & $\psi\in\mathbb{C}^4$ & $\psi\in\mathbb{H}^2\otimes_\mathbb{R}\mathbb{C}$ \\
        Pauli Matrices & $\vec{\sigma}$ & $\Sigma_\ell=\mathbb{R}\{ih,ij,ik\}$ \\
         Dirac Matrices  & $\gamma^\mu=\begin{pmatrix}
0 & \sigma^\mu \\
\tilde{\sigma}^\mu & 0 
\end{pmatrix}$ & $\Gamma^\mu=\begin{pmatrix}
0 & \Sigma^\mu \\
\tilde{\Sigma}^\mu & 0 
\end{pmatrix}$ \\
         Helicity & $ \frac{1}{2\sqrt{\vec{p}\cdot\vec{p}}} \begin{pmatrix}
    \vec{\sigma}\cdot\vec{p} & 0 \\
    0 & \vec{\sigma}\cdot\vec{p}
\end{pmatrix}$  & $\frac{1}{2\sqrt{p_\ell p^\ell}} \begin{pmatrix}
    i\omega_\ell p_\ell & 0 \\
    0 & i\omega_\ell p_\ell
\end{pmatrix}$ \\
         Magnetic Moment & $\vec{\mu}=\frac{e}{2m}\vec{\sigma}$ & $\mu_\ell=\frac{e}{2m}\Sigma_\ell$\\
         U(1) Conserved Current & $\bar{\psi}\gamma^\mu\psi$ & $\bar{\psi}\Gamma^\mu\psi$\\
         Weak Isospin and Hypercharge & $\mathfrak{su}(2)_L\oplus\mathfrak{u}(1)_Y$ & $\mathfrak{su}(2)_L\oplus\mathfrak{u}(1)_Y$\\
         \hline
    \end{tabular}
    \caption{A direct comparison of the standard and complex quaternionic formulations.}
    \label{tab:placeholder}
\end{table}

We noticed that within the algebra $\mathfrak{gl}_2\mathbb{H}\otimes_\mathbb{R}\mathbb{C}$ there is an additional representation of weak isospin and hypercharge that admits similar matrix structure to the standard choices and explored the consequences of SSB.
Several conflicts emerged using this alternative choice. Among these conflicts were a negative definite inner product between the $W^+$ and $W^-$ bosons, and the Z boson mediating a repulsive force. This alternative choice of isospin and hypercharge and its consequences may be useful in theorizing interactions beyond the standard model. 
We may also observe that similarly to how the usual representation of weak isospin is a subalgebra of $\mathfrak{gl}_2\mathbb{H}\otimes_\mathbb{R}\mathbb{C}$, the fundamental representation of $\mathfrak{su}(3)$ is a subalgebra of $\mathfrak{gl}_3\mathbb{H}\otimes_\mathbb{R}\mathbb{C}$. Thus, the spinor formalism developed here is also compatible with the Gell-mann matrices and chromodynamic gauge structure.
The extension to chromodynamics will be explored in future work.

\bibliographystyle{iopart-num}

\bibliography{RefQuaternion}

\end{document}